# Isochronal synchrony and bidirectional communication with delay-coupled nonlinear oscillators

Brian B. Zhou[1] and Rajarshi Roy[1,2,3]

[1]*Institute for Research in Electronics and Applied Physics,*

[2]*Department of Physics, and*

[3]*Institute for Physical Science and Technology,*

University of Maryland, College Park, Maryland 20742

(rroy@glue.umd.edu)

(October 9, 2006)

We propose a basic mechanism for isochronal synchrony and communication with mutually delay-coupled chaotic systems. We show that two Ikeda ring oscillators (IROs), mutually coupled with a propagation delay, synchronize isochronally when both are symmetrically driven by a third Ikeda oscillator. This synchronous operation, unstable in the two delay-coupled oscillators alone, facilitates simultaneous, bidirectional communication of messages with chaotic carrier waveforms. This approach to combine both bidirectional and unidirectional coupling represents an application of generalized synchronization using a mediating drive signal for a spatially distributed and internally synchronized multi-component system.

05.45.Vx, 05.45.Xt, 42.65.Sf

A particularly striking realization in the dynamics of coupled oscillators is that even systems displaying chaotic waveforms will synchronize when appropriately coupled.[1][2] Chaos synchronization has attracted wide-spread research interest from both scientists and engineers by providing insights into natural phenomena and motivation for practical applications in communications and control.[3][4][5] Recently, high speed, long distance transmission of messages using synchronized chaotic lasers was demonstrated in a commercial fiber-optic network.[6]

Traditionally, coupling schemes between nonlinear oscillators have been classified as either unidirectional or bidirectional. In Refs. [1,2], Pecora and Carroll demonstrated chaos synchronization in unidirectionally coupled systems, where one nonlinear oscillator influences the dynamics of another but not conversely. Most chaos-based communication techniques in electronic[7] and optical[8] systems, including that used in [6], leverage synchrony in unidirectional *drive-response* systems. However, a fundamental limitation of these schemes is that messages may be passed in only one direction due to the clear distinction between transmitter (drive) and receiver (response). In this paper, we inquire into the feasibility of *bidirectional* chaos communication through a single system. Simultaneous, two-way transmission of messages logically compels bidirectional coupling between the communicating systems. The dynamics of each mutually coupled oscillator are hence interdependent, promoting a potential transmitter-receiver duality. However, delays in the coupling interaction, arising from the finite speed of signal transmission, can no longer be ignored as they are in unidirectional systems. They now pose fundamental challenges to synchronization.

When two identical nonlinear chaotic oscillators are bidirectionally coupled with a time delay for the propagation of signals between them, it is well known that they do not display stable isochronal synchrony. Any slight asymmetry between the two systems, such as different initial states or experimental noise, will prevent the isochronal solution. One realistic possibility is that the two oscillators synchronize instead with a time delay given by the propagation time (achronal synchrony),[9][10] a behavior characteristic of unidirectionally delay-coupled systems. However, achronal synchrony in bidirectional systems is both less stable and exact, making it non-ideal for use in communication. For example, the roles of leader and follower often switch randomly between the two oscillators.[11] Sometimes, the signal from *either* system can be shifted by the time delay to reveal approximate synchrony, in which case no clear leader or follower can be defined.[12] Finally, achronal synchronization errors are only vanishing for periodic signals.

We report here a method to achieve stable isochronal synchrony between two mutually delay-coupled oscillators through use of a third dynamical system. Isochronal synchrony is important for practical purposes as it enables symmetric protocols in bidirectional communication and sidesteps the instabilities in delay-coupled systems discussed above. In Ref. [13], isochronal synchrony between two mutually delay-coupled semiconductor lasers was attained by adding to each laser self-feedback loops matched to the coupling delay time.[13] Moreover, Fischer and colleagues[14] recently studied isochronal synchrony between the outer oscillators in a chain of three mutually delay-coupled oscillators, extending the results of an earlier experiment where coupling was instantaneous.[15] We implement a variation based on generalized synchronization that combines bidirectional and unidirectional coupling and is both robust and direct. The model nonlinear system of our study is the Ikeda ring oscillator (IRO), a simplified representation of a ring laser. As depicted in Fig. 1, IROs 1 and 2 are mutually coupled to each other with delays corresponding to travel time in passive fiber. IRO3 is unidirectionally coupled to them both. Hence, IRO3 influences the waveforms of the mutually coupled Ikeda oscillators, but not vice versa; indeed these latter oscillators are generally synchronized to IRO3. If no transmitted message is present, we show that isochronal synchrony will result between IROs 1 and 2. We have indicated schematically how independent messages may be simultaneously encoded in the waveforms generated by IROs 1 and 2, transmitted through the communication channel, and recovered at the opposite oscillator.

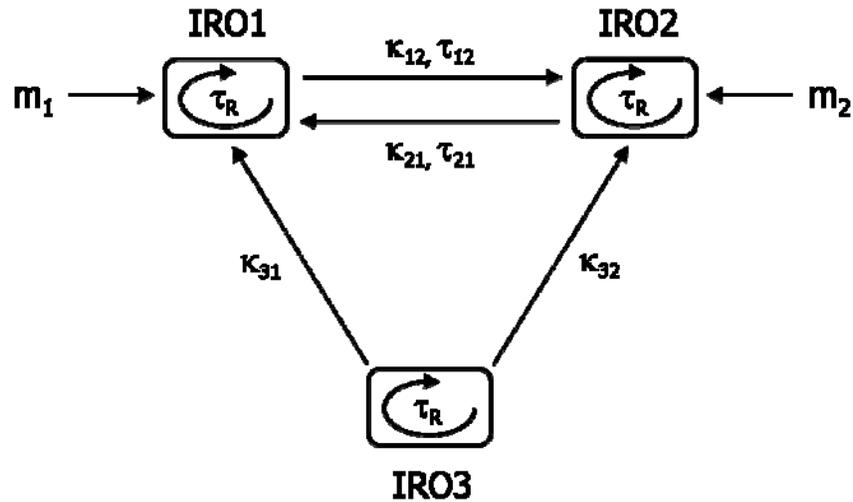

**FIG. 1**. Scheme for isochronal synchrony. IROs 1 and 2 are bidirectionally coupled, while IRO3 drives them both. In bidirectional communication, messages

1 ($m_1$) and 2 ($m_2$) are injected at IROs 1 and 2, respectively. $\tau_R$ denotes the ring roundtrip time of each IRO.

Abarbanel and Kennel[16] introduced the Ikeda ring oscillator as a basic, but characteristic model for fiber ring lasers. Each oscillator is itself a nonlinear system with feedback delayed by the roundtrip time $\tau_R$. Reference [16] demonstrated that such Ikeda systems could generate high dimensional waveforms and be well synchronized for unidirectional coupling. A message waveform injected into the transmitter could be masked and successfully recovered at the receiver through subtraction of its input and output signals. No propagation delay was considered in the communication channel, since the coupling was unidirectional. There was no attempt to communicate information bidirectionally between the two systems at the same time.

For a single free-running IRO, the dynamics at a fixed point in the ring cavity are described by a discrete time map for the complex electric field envelope $E(t)$ and an ordinary differential equation for the spatially averaged population inversion $w(t)$:

$$E(t+\tau_R) = E_I\, e^{i(\omega_I - \omega_0)t} + B\, e^{i\phi}\, E(t)\, e^{(\beta + i\alpha)w(t)} \qquad (1)$$

$$\frac{dw(t)}{dt} = Q - 2\gamma\{w(t) + 1 + |E(t)|^2 (e^{Gw(t)} - 1)/G\}. \qquad (2)$$

Similarly to Ref. [16], we choose the injected field $E_I = 1$, detuning $\omega_I - \omega_0 = 0$, return coefficient $B = 0.8$, propagation phase change $\phi = 0.4$, and gain parameters, $\beta = 0$, $\alpha = 6$, $G = 0.01$. Additionally, we set the pumping $Q = 0$ and the normalized atomic decay rate $\gamma = 1$. Time is normalized to units of the roundtrip time $\tau_R$, which we take to be unity. Hence, oscillations are sustained by the injected field $E_I$ rather than by conventional pump and amplification. Our parameter choices aim to most transparently demonstrate synchrony and communication through our coupling arrangement, not to model exact experimental conditions. Advanced discussion of the model, including experimentally accurate parameter values, can be found in Refs. [17-19].[17][18][19]

The three IRO coupling scheme of Fig. 1 is numerically represented by substituting alternate forms $E'(t)$ for $E(t)$ in the governing dynamical equations. We may write:

$$E_{1,2,3}(t+1) = f[E'_{1,2,3}(t), w_{1,2,3}(t)] \qquad (3)$$

$$\frac{dw_{1,2,3}(t)}{dt} = g[w_{1,2,3}(t), |E'_{1,2,3}(t)|^2], \qquad (4)$$

where

$$E'_1(t) = (1 - \kappa_{12} - \kappa_{31}) E_1(t) + \kappa_{21} E_2(t - \tau_{21}) + \kappa_{31} E_3(t) \qquad (5)$$

$$E'_2(t) = (1 - \kappa_{21} - \kappa_{32}) E_2(t) + \kappa_{12} E_1(t - \tau_{12}) + \kappa_{32} E_3(t) \qquad (6)$$

$$E'_3(t) = E_3(t). \qquad (7)$$

For simplicity, propagation delays from the drive system IRO3 to either of the mutually coupled IROs are ignored. Our results hold as long as the unidirectional delays from IRO3 are equal; that is, IRO3 symmetrically influences IROs 1 and 2. We further assume that the time delay $\tau_{12}$ of the signal from IRO1 into IRO2 and the delay $\tau_{21}$ from IRO2 to IRO1 are both equal to 3.14. These mutual coupling delays would certainly be matched if a single communication channel connected IROs 1 and 2. We choose the mutual coupling strengths $\kappa_{12} = \kappa_{21} = 0.3$ ($\kappa_{12}$) and the drive coupling strengths $\kappa_{31} = \kappa_{32} = 0.4$ ($\kappa_{3}$). Though the drive signal is unidirectionally injected, we subtract an equivalent fraction from the self-feedback field of IROs 1 and 2 in order to match power as $\kappa_3$ is varied. This maintains operation of all Ikeda oscillators in the same basin of attraction, facilitating comparisons of their waveforms. We integrate equations (3) and (4) using a four-order Runge-Kutta routine with step size 0.02.

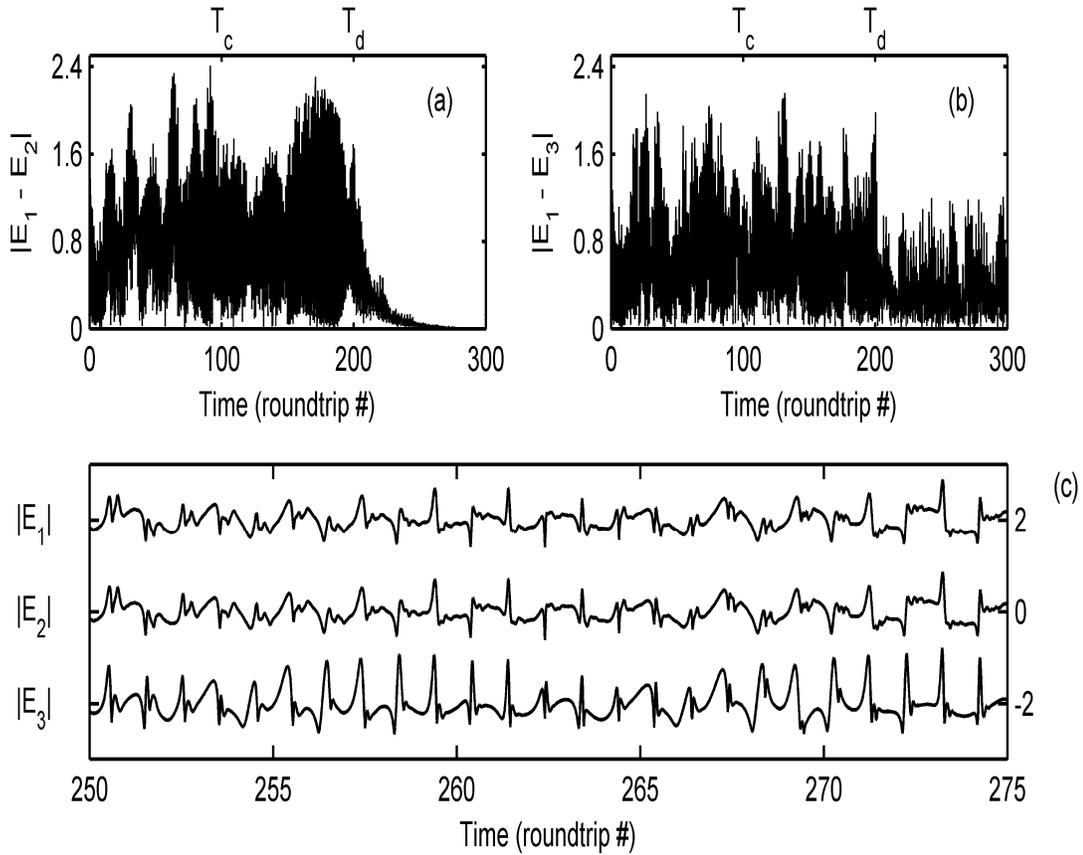

**FIG. 2**. Synchronization errors (a) $|E_1 - E_2|$ between the mutually coupled oscillators and (b) $|E_1 - E_3|$ between IRO1 and drive system IRO3. When only mutual coupling is initiated at time $T_c = 100$, no synchronization between IROs 1 and 2 is observed. After injection of IRO3's signal at $T_d = 200$, IROs 1 and 2 synchronize isochronally, but remain different from the drive. (c) A close-up of the three waveforms after synchronization. The mean values of $|E_1|$, $|E_2|$, and $|E_3|$ are offset to 2, 0, and -2, respectively.

We show in Fig. 2(a) that the waveforms of mutually coupled IROs 1 and 2 synchronize isochronally to each other after unidirectional coupling with driver IRO3 is initiated at time $T_d$. Before $T_d$, there is no sustained synchrony (isochronal or achronal) between IROs 1 and 2. After $T_d$, the difference between the waveforms of IROs 1 and 2 decreases rapidly. In Fig. 2(c), we display without time-shift a twenty-five roundtrip zoom of the three waveforms after synchrony is achieved. It is evident that the synchronized waveforms remain quite different from the driving waveform of IRO3 (see also Fig. 2(b)). In Fig. 3, we plot the isochronal cross-correlation $\rho_{ij}$ between the intensity time traces of IROs $i$ and $j$ ($I_i = |E_i|^2$) as the strength of the drive $\kappa_3$ is varied, holding constant $\kappa_{12} = 0.3$. We calculate $\rho_{ij}$, given by

$$\rho_{ij} = \frac{\sum_n (I_{i,n} - \langle I_i \rangle)(I_{j,n} - \langle I_j \rangle)}{\sqrt{\sum_n (I_{i,n} - \langle I_i \rangle)^2 \sum_n (I_{j,n} - \langle I_j \rangle)^2}}, \qquad (8)$$

for time series of five-hundred roundtrips after waiting five-hundred roundtrips for transients to decay. The subscript $n$ for the summation refers to the time step, and $\langle ... \rangle$ denotes the mean value over the entire time series.

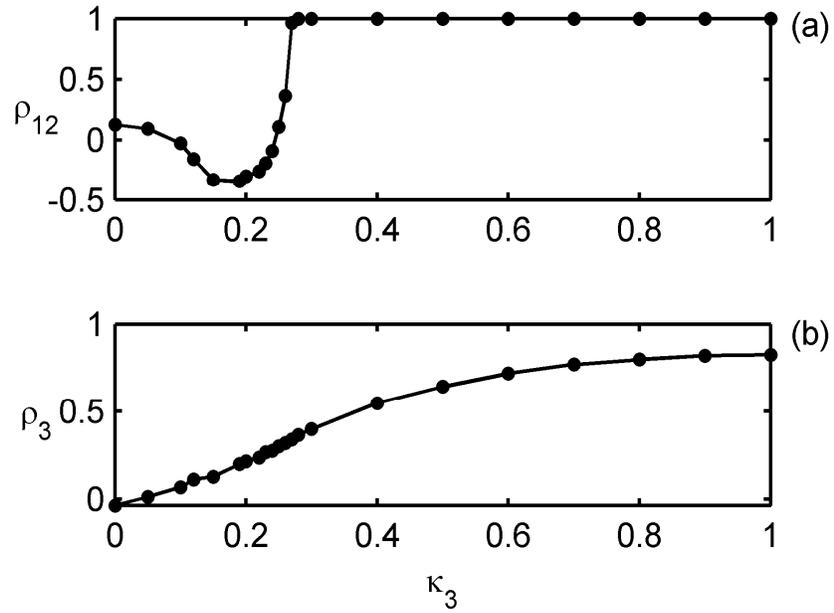

**FIG. 3.** The cross correlations (a) $\rho_{12}$ between IROs 1 and 2 and (b) $\rho_3 = (\rho_{31} + \rho_{32})/2$, the average cross correlation between IROs 1 and 2 to drive system IRO3, as a function of the drive coupling strength $\kappa_3$, with $\kappa_{12} = 0.3$ constant. Isochronal synchrony occurs above critical coupling strength $\kappa_3^* \approx 0.27$.

We observe isochronal synchrony between IROs 1 and 2 above a critical drive strength $\kappa_3^* \approx 0.27$. As a result of the nonzero bidirectional coupling, the driven waveforms of IROs 1 and 2 synchronize identically to each other, but not to the drive, even though the drive system is perfectly matched. This is evidenced by the average correlation of the driven waveforms to the drive signal $\rho_3 = (\rho_{31} + \rho_{32})/2$, which increases with the drive strength to a limiting value that is less than one (complete synchrony). For private communications, the optimal setting of the drive strength should be just above $\kappa_3^*$ in order to minimize the resemblance of the carrier waveforms to the drive waveform, yet maintain synchrony between the communicating systems.

The dynamics of our coupling scheme is in essence the generalized synchronization (GS)[20] of a spatially distributed system (IROs 1 and 2) to the driver IRO3. The drive signal alone, rather than the initial conditions of IROs 1 and 2, determines the long-run synchronized behavior. We emphasize that the significant difference here from the usual situation in GS is that the driven "system" is actually composed of mutually coupled subsystems, which identically synchronize. In addition, this synchrony between the mutually coupled oscillators is not contingent on the dynamical nature of the drive signal. In Fig. 4, we have synchronized IROs 1 and 2 using as the drive system (a) an IRO that is highly parameter mismatched and (b) a chaotic Rössler oscillator, whose low dimensionality contrasts sharply with the driven Ikeda systems. The Rössler equations[21] used were

$$\dot{x} = \lambda(-y-z)$$

$$\dot{y} = \lambda(x+ay) \qquad (9)$$

$$\dot{z} = \lambda(b+z(x-c))$$

where $a = b = 0.2$, $c = 5.7$, and $\lambda = 20$. The $x$ and $y$ components of the generated Rössler time series was then taken to be the real and imaginary parts, respectively, of the complex driving waveform. The mean amplitude of this drive was scaled to be 0.5 by a constant multiplicative factor. Coupling coefficients for Fig. 4 were $\kappa_{12} = 0.3$ and $\kappa_3 = 0.4$.

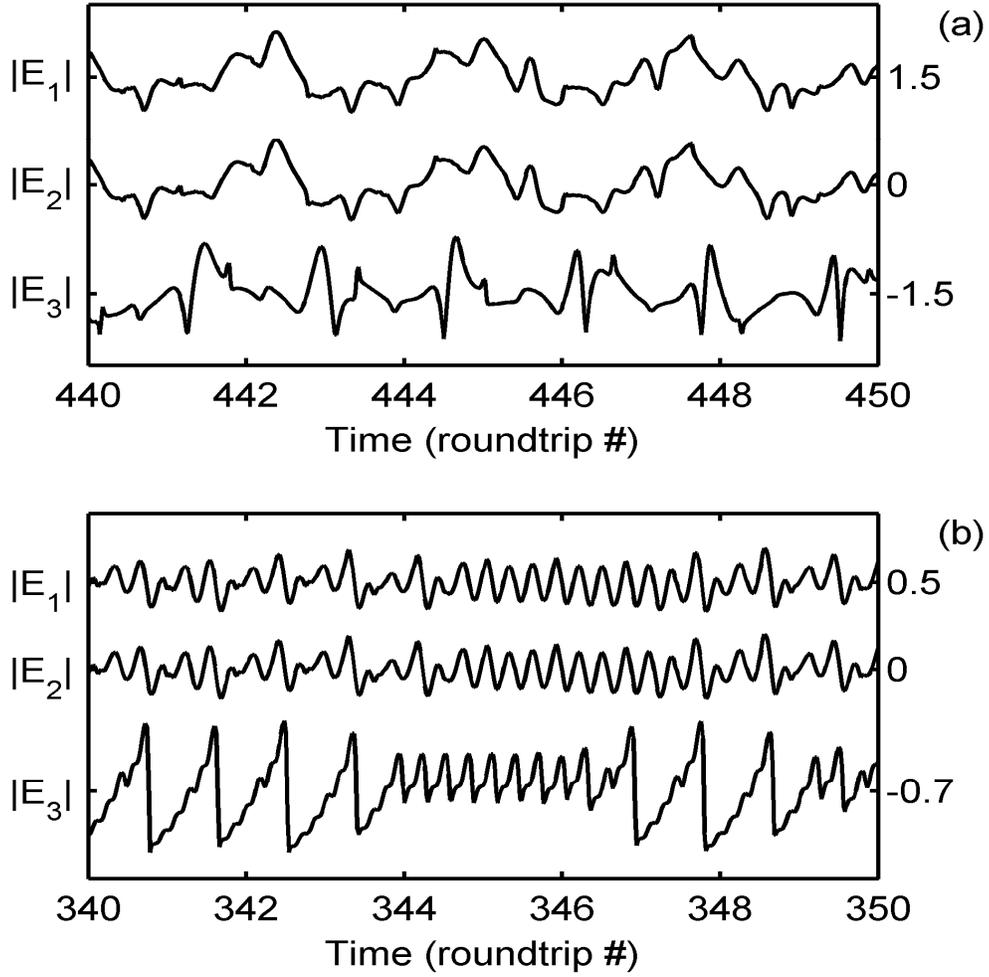

**FIG 4.** Time traces when the drive system IRO3 is (a) a parameter mismatched Ikeda oscillator ($\tau_R = 1.62$, $E_I = 1.05$, $B = 0.85$, $\beta = 0.01$, $\alpha = 5.5$, $G = 0.02$, $\gamma = 0.8$) and (b) a Rössler oscillator. In both cases, IROs 1 and 2 maintain isochronal synchrony while adopting qualitatively the character of the drive signal. Mean values are offset for display.

Perhaps a more intuitive interpretation of our scheme comes from comparison with the auxiliary systems approach[22] in detecting GS. GS implies that the dynamical evolution of the mutually coupled systems depends solely on the drive signal; hence, it should not matter in the long run whether we initiate first mutual coupling between IROs 1 and 2 and then coupling to the drive signal (as we do in our simulations) or if we reverse that order. The situation of driving both IROs 1 and 2 without bidirectional coupling between them is identical to that of the auxiliary systems approach. After IROs 1 and 2 identically synchronize, indicating GS, turning on bidirectional coupling with matched delays will not perturb the isochronal synchrony since the action is symmetric. However, the final GS state, independent of the steps that establish it, now incorporates three separate time scales: the intrinsic time scales of the drive and response oscillators and the delay time between the response oscillators.

With isochronal synchrony established, we show that it is possible to communicate simultaneously in a bidirectional fashion between IROs 1 and 2. However, straightforward application of the unidirectional chaos modulation technique[23] to each communicating system no longer preserves synchrony. Under this first encoding process, the input fields to the dynamical equations for IROs 1 and 2, denoted now by $E''_{1,2}(t)$, become:

$$E''_1(t) = E'_1(t) + m_1(t) + m_2(t - \tau_{21}) \qquad (10)$$

$$E''_2(t) = E'_2(t) + m_2(t) + m_1(t - \tau_{12}) \qquad (11)$$

where $E'_{1,2}(t)$ retain their previous forms. The two additional message terms correspond to simultaneously injecting the transmitted message both into the transmitting ring cavity (i.e., $m_1(t)$ in (10)) and into the coupling line, reaching the opposite oscillator after travel to give the received message term (i.e., $m_2(t-\tau_{21})$ in (10)). Clearly, the difference $m_1(t) - m_2(t)$ perturbs the established synchrony, and any successful decoding occurs in spite of the messages. For bidirectional communication, we instead adapt the traditional chaos modulation technique:

$$E_1''(t) = E_1'(t) + m_1(t-\tau_{12}) + m_2(t-\tau_{21}) \qquad (12)$$

$$E_2''(t) = E_2'(t) + m_2(t-\tau_{21}) + m_1(t-\tau_{12}), \qquad (13)$$

We now inject the transmitted message, *delayed by the propagation time*, into the transmitting ring cavity. This encoding process for bidirectional communication, while more involved than that used in unidirectional systems, ensures that the effect of the messages is symmetric, thus preserving complete synchrony. Our scheme retains the key advantages of chaos modulation; transmitted messages may be of arbitrary size and actively influence the dynamical evolution.

Since $E_1(t)$ synchronizes identically to $E_2(t)$, decoding (at time *t*) is accomplished by subtracting the receiver's own cavity field, delayed by the propagation time, from the total field received from the opposite oscillator:

$$\kappa_{21} E_2(t-\tau_{21}) + m_2(t-\tau_{21}) - \kappa_{21} E_1(t-\tau_{21}) \to m_2(t-\tau_{21}) \qquad (14)$$

$$\kappa_{12} E_1(t-\tau_{12}) + m_1(t-\tau_{12}) - \kappa_{12} E_2(t-\tau_{12}) \to m_1(t-\tau_{12}). \qquad (15)$$

In Fig. 5(a) and (b), we display the (different) decoded digital messages in the two directions. We have chosen a return-to-zero scheme with message size 0.1 and a bit rate of five random bits per round trip of the Ikeda ring. The top trace of each figure represents the near flawless decoding under idealized assumptions of perfect parameter match between IROs 1 and 2. In the middle trace, we have assumed small parameter mismatches (2%) in $E_I$, $B$, $\alpha$, and $\gamma$. Message fidelity deteriorates, but remains error-free. We note that additive white noise in the field equations has a similar effect on decoding by introducing non-vanishing synchronization errors. Finally, the bottom trace depicts the recovered message using the conventional encoding technique of equations (10) and (11). We note that decoding under this scheme fails completely at higher bit rates and larger message sizes.

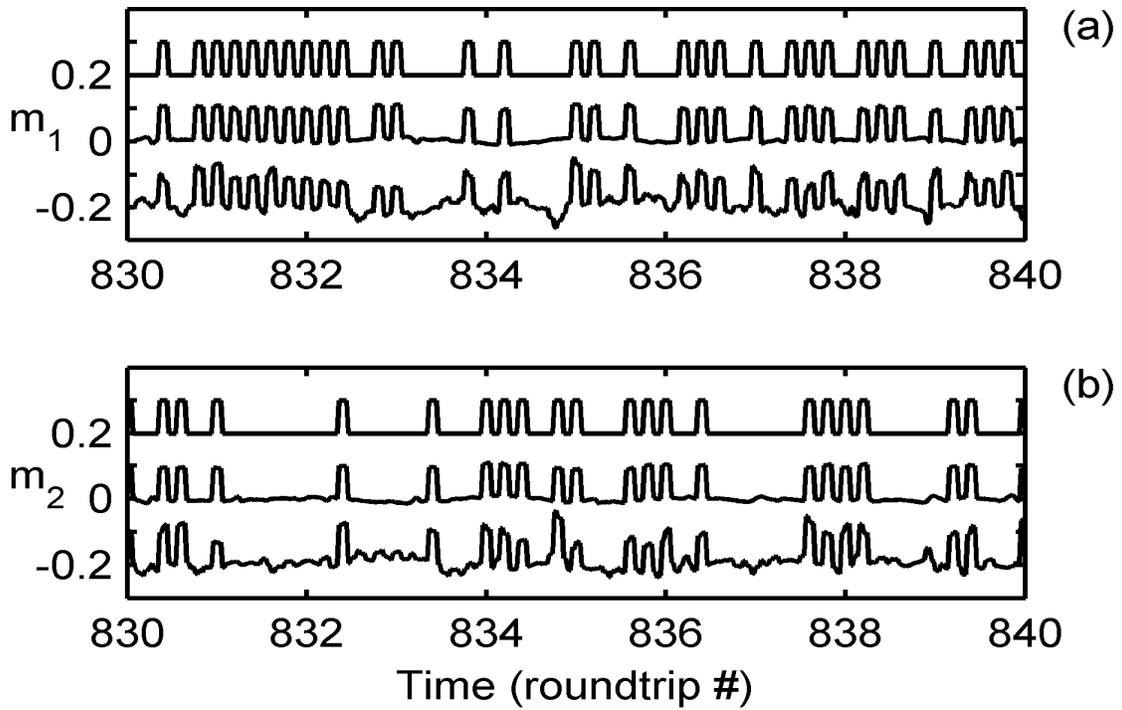

**FIG. 5.** Ten roundtrip excerpts of the recovered messages (a) $m_1(t)$ at IRO2 and (b) $m_2(t)$ at IRO1. The top trace in each figure assumes perfect parameter match between the communicating IROs, while in the middle trace parameters are mismatched by 2% ($E_I = 1.02$, $B = 0.82$, $\alpha = 5.85$, $\gamma = 0.98$ for IRO2). The bottom trace depicts message recovery when injection of the transmitted message into the transmitting cavity is not delayed.

We have tested this scheme for producing isochronal synchrony and bidirectional transmission and recovery of information over a range of coupling coefficients and propagation delays. In addition, bidirectional communication was equally supported by drive sources non-identical to the communicating systems, such as those in Fig 4. The Ikeda model systems used here are representative of systems with an internal time delay coupled together bidirectionally with propagation time delays. We expect that the scheme we have developed will hold in general for delay-coupled dynamical systems under appropriate coupling schemes and for suitable parameter regimes.

In light of the recent interest in developing a public key cryptographic system using chaotic systems,[24] we discuss briefly the security aspect of our scheme while making no claims of immunity against a designed attack. In many unidirectional chaos communication techniques, the evolution of the receiver is determined by a single signal from the transmitter. If the drive signal is intercepted, then the attacker would be able to reconstruct the dynamics of the intended receiver by inputting the intercepted signal into an identical system. However, due to the degeneracy of the synchronized solution, achieving exact parameter match over a potentially large parameter space is extremely difficult. In our bidirectional, three-oscillator scheme, the dynamics of each communicating system are determined by three sources rather than one: the mediating drive signal, the received carrier signal, and *the transmitted message*. Reasonably, the attacker lacks knowledge of the transmitted message. Thus, even with an identical receiver, exact reconstruction of the decoding waveform cannot be performed with the intercepted drive and mutual coupling signals alone. Although in practice attackers can gain significant information on the transmitted messages with only an approximate decoding waveform, we believe further work on our ideas may be able to minimize this threat. Ideally, the transmitted message of each "receiver" would serve the dual function of a private encryption key and be leveraged even when only one message is being sent. Along the same lines, we may view the common drive signal that mediates the synchrony between the two communicating oscillators as the public encryption key.

In summary, we have achieved isochronal synchrony between two bidirectionally, delay-coupled Ikeda ring oscillators through the symmetric injection of a unidirectional signal from an independent chaotic source. The mutually coupled Ikeda oscillators, models for ring lasers, evolve in generalized synchrony with the drive signal under a complex functional relationship that incorporates multiple time scales. We emphasize that this isochronal synchrony, unstable otherwise, is essential for bidirectional information communication. A robust chaos modulation technique is implemented to transmit independent messages simultaneously and bidirectionally. This technique allows both message signals to actively influence the dynamics of the communicating systems, sustaining rather than perturbing synchrony. We have carried out preliminary numerical simulations to test the extension of these ideas to larger numbers of delay-coupled oscillators driven by a common mediator. Experiments to test these concepts and results on systems of fiber ring lasers and semiconductor lasers are in progress.

**Acknowledgements:** We are grateful for valuable conversations with Tony Franz, Nathan Karst, Will Ray, Leah Shaw, and Ira Schwartz. B.Z. acknowledges support from the University of Maryland TREND program. R.R. acknowledges support from the Physics Division of the U.S. Office of Naval Research.

**References:**

[1] L. M. Pecora and T. L. Carroll, Phys. Rev. Lett. **64**, 821-824 (1990).

[2] L. M. Pecora and T. L. Carroll, Phys. Rev. A **44**, 2374-2383 (1991).

[3] A. Pikovsky, M. Rosenblum, and J. Kurths, *Synchronization: A Universal Concept in Nonlinear Science* (Cambridge University Press, Cambridge, U.K., 2001).

[4] S. Boccaletti, J. Kurths, G. Osipov, D. L. Valladares, and C. S. Zhou, Phys. Rep. **366**, 1-101 (2002).

[5] A. Uchida, F. Rogister, J. García-Ojalvo, and R. Roy, Prog. Optics **48**, 203-341 (2005).

[6] A. Argyris, D. Syvridis, L. Larger, V. Annovazzi-Lodi, P. Colet, I. Fischer, J. García-Ojalvo, C. R. Mirasso, L. Pesquera, and K. A. Shore, Nature **438**, 343-346 (2005).

[7] K. M. Cuomo and A. V. Oppenheim, Phys. Rev. Lett. **71**, 65-68 (1993).

[8] G. D. VanWiggeren and Rajarshi Roy, Science **279**, 1198-1200 (1998).

[9] T. Heil, I. Fischer, W. Elsässer, J. Mulet, and C. R. Mirasso, Phys. Rev. Lett. **86**, 795-798 (2001).

[10] J. Mulet, C. Mirasso, T. Heil, and I. Fischer, J. Opt. B: Quantum Semiclass. Opt. **6**, 97-105 (2004).

[11] E. A. Rogers-Dakin, J. García-Ojalvo, D. J. DeShazer, and R. Roy, Phys. Rev. E **73**, 045201 (2006).

[12] L. B. Shaw, I. B. Schwartz, E. A. Rogers, and R. Roy, Chaos **16**, 01511 (2006).

[13] E. Klein, N. Gross, M. Rosenbluh, W. Kinzel, L. Khaykovich, and I. Kanter, Phys. Rev. E **73**, 066214 (2006).

[14] I. Fischer, R. Vicente, J. M. Buldu, M. Peil, C. R. Mirasso, M. C. Torrent, and J. García-Ojalvo, Phys. Rev. Lett. **97,** 123902 (2006).

[15] J. R. Terry, K. S. Thornburg, Jr., D. J. DeShazer, G. D. VanWiggeren, S. Zhu, P. Ashwin, and R. Roy, Phys. Rev. E **59**, 4036-4043 (1999).

[16] H. D. I. Abarbanel and M. B. Kennel, Phys. Rev. Lett. **80**, 3153-3256 (1998).

[17] K. Ikeda, Opt. Commun. **30**, 257-261 (1979).

[18] (a) Q. L. Williams and R. Roy, Opt. Lett. **21**, 1478-1480 (1996); (b) Q.L. Williams, J. García-Ojalvo, and R. Roy, Phys. Rev. A **55**, 2376-2386 (1997).

[19] H. D. I. Abarbanel, M. B. Kennel, M. Buhl, and C. T. Lewis, Phys. Rev. A **60**, 2360-2374 (1999).

[20] N. F. Rulkov, M. M. Sushchik, L. S. Tsimring, and H. D. I. Abarbanel, Phys. Rev. E **51**, 980-994 (1995).

[21] O. E. Rössler, Phys. Lett. **57A**, 397-398 (1976).

[22] H. D. I. Abarbanel, N. F. Rulkov, and M. M. Sushchik, Phys. Rev. E **53**, 4528-4535 (1996).

[23] L. Kocarev and U. Parlitz, Phys. Rev. Lett. **74**, 5028-5031 (1995).

[24] (a) N. Gross, E. Klein, M. Rosenbluh, W. Kinzel. L. Khaykovich, and I Kanter, *arXiv: cond-mat/0507554 v1*, 24 July 2005; (b) E. Klein et al., *arXiv: cond-mat/0604569 v1*, 25 April 2006.